%% file: rdr.tex
\documentclass[aps,prl,twocolumn,showpacs,superscriptaddress,groupedaddress]{revtex4}  
\usepackage{graphicx}  
\usepackage{dcolumn}   
\usepackage{bm}        
\usepackage{amssymb}   
\usepackage{slashed}

\newcommand{\Rdr}{R_{\Delta R}}

\newcommand{\DR}{\Delta R}

\newcommand{\ptnbrmin}{p_{T \rm min}^{\rm nbr}}

\newcommand{\mur}{\mu_R}
\newcommand{\muf}{\mu_F}

\newcommand{\as}{\alpha_s}
\newcommand{\ass}{\alpha_s^2}
\newcommand{\asss}{\alpha_s^3}
\newcommand{\asmz}{\alpha_s(M_Z)}
\newcommand{\asq}{\alpha_s(Q)}
\newcommand{\aspt}{\alpha_s(p_T)}

\newcommand{\pythia}{{\sc pythia}}
\newcommand{\herwig}{{\sc herwig}}
\newcommand{\sherpa}{{\sc sherpa}}

\newcommand{\Rcone}{R_{\rm cone}}

\newcommand{\ord}{{\cal O}}
\newcommand{\ppbar}{p{\bar{p}}}
\newcommand{\epem}{e^+e^-}

\begin{document}

\hspace{5.2in} \mbox{FERMILAB-PUB-12-397-E}

\title{\boldmath
Measurement of angular correlations of jets at $\sqrt{s}=1.96\,$TeV and \\
determination of the strong coupling at high momentum transfers}

\input author_list.tex

\date{September 29, 2012}

\begin{abstract}
We present a measurement of the average value of
a new observable at hadron colliders that is 
sensitive to QCD dynamics and to the strong coupling constant, 
while being only weakly sensitive to parton distribution functions.
The observable measures the angular correlations of jets
and is defined as the number of neighboring jets above a given 
transverse momentum threshold which accompany a given jet
within a given distance $\DR$ in the plane of rapidity and 
azimuthal angle.
The ensemble average over all jets in an inclusive jet sample is measured 
and the results are presented as a function of transverse momentum of the 
inclusive jets, in different regions of $\DR$ and for different
transverse momentum requirements for the neighboring jets.
The measurement is based on a data set corresponding to an 
integrated luminosity of $0.7\,$fb$^{-1}$ collected with the D0 detector
at the Fermilab Tevatron Collider in $\ppbar$ collisions 
at $\sqrt{s}=1.96\,$TeV.
The results are well described by a perturbative QCD calculation
in next-to-leading order in the strong coupling constant,
corrected for non-perturbative effects.
From these results, we extract the strong coupling and test the 
QCD predictions for its running over a range of momentum transfers 
of $50$--$400\,$GeV.
\end{abstract}

\maketitle


Quantum chromodynamics (QCD) predicts that the strong force
between quarks and gluons becomes weaker when probed at 
high momentum transfers, corresponding to small distances.
This property, referred to as asymptotic freedom, is derived from 
the renormalization group equation (RGE)~\cite{Callan:1970yg,RGE2,RGE3}.
The RGE does not predict the value of the strong coupling $\as$,
but it describes the dependence of $\as$ on the renormalization scale $\mur$, 
and therefore on the momentum transfer.
Tests of perturbative QCD (pQCD) and the property of asymptotic freedom
can be divided into tests of the validity of the RGE
and determinations of the value of $\as$.
By convention, $\as$ values extracted from data at different 
momentum transfers are evolved to the common scale 
$\mu_R = M_Z$ to allow comparisons between experiments.
The current world average value is 
$\asmz = 0.1184\pm 0.0007$~\cite{PDG2012}.
The validity of the RGE is tested by studying the dependence
of $\as$ on  the momentum transfer.
At present, the RGE predictions have been tested in 
deep-inelastic $e^{\pm}p$ scattering and in
$\epem$ annihilation,
where $\as$ results have been obtained for momentum transfers 
up to $208\,$GeV~\cite{PDG2012}.
Attempts to extract $\as$ at higher momentum transfers 
have been carried out using inclusive jet cross section data in 
hadron-hadron collisions~\cite{Affolder:2001hn,atlas}.
These analysis methods require parton distribution functions (PDFs) 
of the proton at large scales as input.
Since the main constraints on PDFs come from data at lower scales, the 
knowledge of PDFs at large scales is mainly based on the evolution according 
to the Dokshitzer-Gribov-Lipatov-Altarelli-Parisi (DGLAP) evolution 
equations~\cite{Gribov:1972ri,Altarelli:1977zs,Dokshitzer:1977sg}
which use $\as$ and the RGE as input.
The $\as$ results from inclusive jet cross section data at high momentum 
transfers can therefore not be regarded as tests of the RGE, since they 
are obtained assuming its validity.

In this Letter a new observable for hadron-hadron collisions
is introduced and its average value is measured. 
It is related to the angular correlations of jets.
In pQCD, this quantity is computed as a ratio of 
jet cross sections, which is proportional to $\as$.
Since PDF dependencies largely cancel in the ratio, the extracted $\as$ 
results are almost independent of initial assumptions on the RGE.
Values of $\as$ are extracted for momentum transfers between $50$ 
and $400\,$GeV.
These provide the first test of the RGE at momentum transfers 
above 208\,GeV.


The analysis presented in this Letter studies the properties of
multi-jet production based on an inclusive jet sample in $\ppbar$ 
collisions at $\sqrt{s}=1.96\,$TeV.
While pQCD predictions for any cross section at a hadron collider
depend on the PDFs, quantities with significantly reduced PDF sensitivity
can be constructed.
One class of such quantities is ratios of three-jet and dijet
cross sections.
Based on such ratios, one can exploit the high energy reach at hadron 
colliders to determine $\as$ and to test the predictions of the RGE 
at previously unexplored momentum scales.
A new observable is introduced, which probes the angular correlations 
of jets in the plane of rapidity $y$~\cite{rapidity} 
and azimuthal angle $\phi$.
This observable measures the number of neighboring jets that accompany 
a given jet with transverse momentum ($p_T$) with respect to the beam axis.
The measured quantity $\Rdr$ is the ensemble average over all jets 
in an inclusive jet sample of this observable.
The inclusive jet sample consists of all jets in a given data set, and 
these jets are hereafter referred to as ``inclusive jets''.
The measured quantity is given by
\begin{equation}
  \Rdr(p_T,\DR,\ptnbrmin) =  
  \frac{\sum_{i=1}^{N_{\rm jet}(p_T)} N_{\rm nbr}^{(i)}(\DR,\ptnbrmin)}%
  {N_{\rm jet}(p_T)}              \label{eq:rdr}
\end{equation}
where $N_{\rm jet}(p_T)$ is the number of inclusive jets
in a given inclusive jet $p_T$ bin, and
$N_{\rm nbr}^{(i)}(\DR,\ptnbrmin)$ is the number of neighboring jets 
with transverse momenta greater than $\ptnbrmin$,
separated from the $i$-th inclusive jet by a distance $\DR$
within a specified interval $\DR_{\rm min}< \DR <\DR_{\rm max}$
with $\DR \equiv \sqrt{(\Delta y)^2 + (\Delta \phi)^2}$.
For $\DR < \pi$, only topologies with at least three jets 
contribute to the numerator of Eq.~(\ref{eq:rdr}), in pQCD, 
and $\Rdr$ is computed at lowest order as a ratio 
of three-jet ($\ord(\asss)$) and inclusive jet cross sections ($\ord(\ass))$.
This ratio is proportional to $\as$.

This measurement is based on a data set corresponding to an 
integrated luminosity of $0.7\,$fb$^{-1}$ collected with the D0 detector
at the Fermilab Tevatron Collider.
$\Rdr(p_T, \DR, \ptnbrmin)$ is measured in an inclusive jet sample
at central rapidities $|y|<1$ for $p_T > 50\,$GeV,
defined by the Run~II midpoint cone jet algorithm~\cite{run2cone} 
with a cone of radius $\Rcone =0.7$ in $y$ and $\phi$. 
It is measured triple differentially,
as a function of inclusive jet $p_T$, 
for different $\ptnbrmin$, and in different $\DR$ regions.
The $\ptnbrmin$ requirements are 30, 50, 70, or 90\,GeV, respectively,
and the different $\DR$ intervals are
$1.4<\DR<1.8$, $1.8<\DR<2.2$, and $2.2<\DR<2.6$.
For jets with $\Rcone = 0.7$, the lower limit of $\DR > 1.4$ ensures
that a jet does not overlap with its neighboring jets.
The upper limit on $\DR$ is smaller than $\pi$, so that contributing 
neighboring jets stem only from three- (or more) jet topologies.
The lowest $\ptnbrmin$ requirement is chosen to ensure that the jet energy 
calibration and the jet $p_T$ resolutions are well understood.
The trigger efficiencies are high for jets with $p_T > 50$\,GeV
in the inclusive jet sample. 
The requirement of $|y|<1$ implies that $(|y| + \DR) < 3.6$ 
over the whole analysis phase space.
In this rapidity region jets are well-measured in the D0 detector.
The data are corrected for experimental effects and are presented 
at the ``particle level'', which includes all stable particles 
as defined in Ref.~\cite{Buttar:2008jx}.

\begin{figure*}
\includegraphics[scale=1.0]{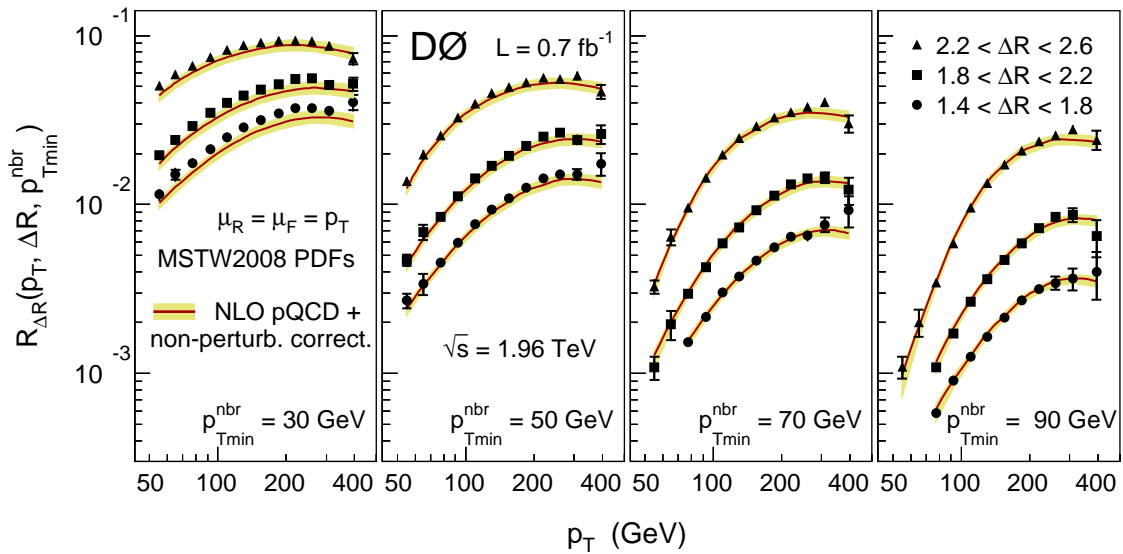}
\caption{(Color online.)
The measurement of $\Rdr$ as a function of inclusive jet $p_T$
for three different intervals in $\DR$ and for four different 
requirements of $\ptnbrmin$.
The inner uncertainty bars indicate the statistical uncertainties,
and the total uncertainty bars display the quadratic sum of 
the statistical and systematic uncertainties.
The theory predictions are shown with their uncertainties.
\label{fig:fig1} }
\end{figure*}


A detailed description of the D0 detector can be found in
Ref.~\cite{d0det}.
The event selection, jet reconstruction, and jet energy and momentum 
correction follow closely those used in recent D0 measurements of 
inclusive jet, dijet and three-jet production
rates~\cite{:2008hua,Abazov:2011vi,:2009mh,Abazov:2010fr,Abazov:2011ub}.
Jets are reconstructed in the finely segmented liquid-argon/uranium 
calorimeter which covers most of the solid angle for polar angles of
$1.7^\circ \lesssim \theta \lesssim 178.3^\circ$~\cite{d0det}.
For this measurement, events are triggered by jet triggers.
Trigger efficiencies are studied as a function of jet $p_T$
by comparing the inclusive jet cross section in data sets obtained by 
triggers with different $p_T$ thresholds in regions where the trigger 
with lower threshold is fully efficient. 
The trigger with lowest $p_T$ threshold is shown to be fully efficient
by studying an event sample obtained independently with a muon trigger.
In each inclusive jet $p_T$ bin, events are taken from a single trigger
which has an efficiency higher than 99\%.

The position of the $p\bar{p}$ interaction is determined from 
the tracks reconstructed using data from the silicon detector and 
scintillating fiber tracker located inside a $2\,$T 
solenoidal magnet~\cite{d0det}.
The position is required to be within $50$\,cm of the detector center
in the coordinate along the beam axis, with at least three tracks 
pointing to it.
These requirements discard (7--9)\% of the events, depending on the 
trigger used.
Contributions from cosmic ray events are suppressed by requiring the 
missing transverse momentum in an event to be less than 70\% (50\%) 
of the uncorrected leading jet $p_T$ if the latter is below (above) 100\,GeV.
The efficiency of this requirement for signal is found to 
be $>99.5\%$~\cite{:2008hua,Abazov:2011vi}.
Requirements on the characteristics of calorimeter shower shapes are
used to suppress the remaining background due to electrons, photons, 
and detector noise that would otherwise mimic jets. 
The efficiency for the shower shape requirements is above $97.5\%$, 
and the fraction of background events is below $0.1$\% for all $p_T$,
as determined from distributions in signal and 
in background-enriched event samples.

The jet four-momenta reconstructed from calorimeter energy depositions
are then corrected, on average, for the response of the calorimeter, 
the net energy flow through the jet cone,
additional energy from previous beam crossings, and 
multiple $p\bar{p}$ interactions in the same event, but not for 
muons and neutrinos~\cite{:2008hua,Abazov:2011vi,Voutilainen2008}.
The absolute energy calibration is determined from 
$Z \rightarrow \epem$ events and the 
$p_T$ imbalance in $\gamma$ + jet events in the region $|y| < 0.4$.
The extension to larger rapidities is derived from dijet events
using a similar data-driven method.
In addition, corrections in the range (2--4)\% are applied that take
into account the difference in calorimeter response due to the
difference in the fractional contributions of quark and 
gluon-initiated jets in the dijet and the $\gamma$ + jet event samples.
These corrections are determined using jets simulated
with the \pythia\ event generator~\cite{pythia} that have been 
passed through a {\sc geant}-based detector simulation~\cite{geant}.
The total corrections of the jet four-momenta vary between 50\% and 
20\% for jet $p_T$ between 50 and 400\,GeV.
An additional correction is applied for systematic shifts in $|y|$ 
due to detector effects~\cite{:2008hua,Abazov:2011vi}.
These corrections adjust the reconstructed jet energy to the
energy of the stable particles that enter the
calorimeter except for muons and neutrinos.

\begin{figure*}
\includegraphics[scale=0.97]{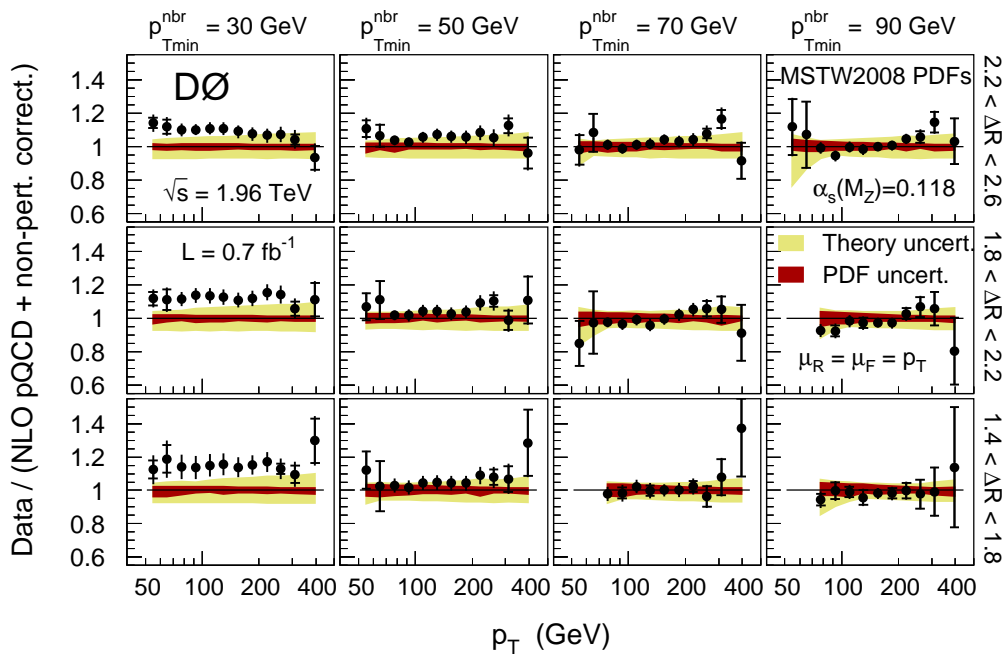}
\caption{(Color online.) 
The ratios of the $\Rdr$ measurements and the theory predictions
obtained for MSTW2008NLO PDFs and $\asmz=0.118$.
The ratios are shown as a function of inclusive jet $p_T$ 
in different regions of $\DR$ (rows)
and for different $\ptnbrmin$ requirements (columns).
The inner uncertainty bars indicate the statistical uncertainties,
and the total uncertainty bars display the quadratic sum of 
the statistical and systematic uncertainties.
The theory uncertainty is the quadratic sum of PDF and scale uncertainties.
\label{fig:fig2}}
\end{figure*}


The differential distributions $\Rdr(p_T, \DR, \ptnbrmin)$
are corrected for experimental effects.
Particle-level events are generated with \sherpa\ 1.1.3~\cite{sherpa}
with MSTW2008LO PDFs~\cite{Martin:2009iq} and with \pythia\ 6.419~\cite{pythia}
with CTEQ6.6 PDFs~\cite{Nadolsky:2008zw} and tune QW~\cite{Group:2006rt}.
The jets from these events are processed by a fast simulation 
of the D0 detector response.
The simulation is based on parameterizations of jet $p_T$ resolutions and 
jet reconstruction efficiencies determined from data and of resolutions 
of the polar and azimuthal angles of jets, which are obtained from a 
detailed simulation of the detector using {\sc geant}.

The $p_T$ resolution for jets is about 15\% at 40 GeV, decreasing 
to less than 10\% at 400 GeV.
To use the fast simulation to correct for experimental effects, the 
simulation must describe all relevant distributions, including the 
$p_T$, $y$ and $\DR$ distributions for the inclusive jets and the 
neighboring jets.
The generated events are reweighted, based on the properties of the 
generated jets, to match these distributions in data.
To minimize migrations between inclusive jet $p_T$ bins due to resolution 
effects, we use the simulation to obtain a rescaling function in 
reconstructed $p_T$ that optimizes the correlation between the 
reconstructed and true values.
The bin sizes in the $p_T$ distributions are chosen to be approximately 
twice the $p_T$ resolution.
The bin purity after $p_T$ rescaling, defined as the fraction of all 
reconstructed events that were generated in the same bin, is above 50\% 
for all bins.
We then use the simulation to determine bin correction factors for 
experimental effects for all analysis bins.
The correction factors are computed bin-by-bin as the ratio of 
$\Rdr$ without and with simulation of the detector response.
These also include corrections for the energies of unreconstructed
muons and neutrinos inside the jets.
The total correction factors for $\Rdr$ using the reweighted \pythia\ 
and \sherpa\ simulations agree typically within 2\%.
The average factors, used to correct the data, are typically between 
0.98 and 1.01, but never below 0.93 or above 1.03.
The difference between the average and the individual corrections is 
taken into account as an uncertainty which is split into two contributions.
One contribution corresponds to the systematic difference between the
two individual corrections, and the other one corresponds to the 
statistical fluctuations.
The former is attributed to the model dependence and assumed to be 
correlated between the data points, while the latter is included in the 
statistical uncertainty of the results.

In total, 69 independent sources of experimental systematic uncertainties
are identified, mostly related to jet energy calibration and jet $p_T$ 
resolution.
The effects of each source are taken as fully correlated between all 
data points.
The dominant uncertainties for the differential cross sections are due 
to the jet energy calibration (2--5)\%, and the model dependence of the 
correction factors (2--3)\%.
Smaller contributions come from the jet $p_T$ resolution (0.5--1.5)\%,
the jet $\phi$ resolution (0.5--2)\%, and from the uncertainties in 
systematic shifts in $y$ (0.5--1)\%.
All other sources are negligible.
The total systematic uncertainties are between 2\% and 6\%.


The results for  $\Rdr(p_T, \DR, \ptnbrmin)$ are displayed in 
Fig.~\ref{fig:fig1} as a function of inclusive jet $p_T$, in different 
regions of $\DR$ and for different $\ptnbrmin$.
The values of $p_T$ at which the data points are presented correspond 
to the geometric bin centers.
A detailed documentation of the results, including the individual 
uncertainty contributions, is provided in the supplementary 
material.  
For a given $\DR$ region, and $\ptnbrmin$, $\Rdr$ increases with $p_T$ 
up to a maximum value, above which it falls when approaching the kinematic 
limit.
At fixed $p_T$, $\Rdr$ increases with $\DR$ and decreases with increasing 
$\ptnbrmin$.
At lower $p_T$, $\Rdr$ depends more strongly on $\ptnbrmin$.
For larger $\ptnbrmin$, both the $p_T$ and the $\DR$ dependencies are stronger.


The theory predictions for $\Rdr$ which are compared to the data, 
and which are later used to extract $\as$, are given by the product 
of the NLO pQCD results and correction factors for non-perturbative 
effects, including hadronization and underlying event.
The non-perturbative corrections are determined using \pythia\ 6.425 with 
tunes AMBT1~\cite{Diehl:2010zz} and DW~\cite{Albrow:2006rt}, 
which use different parton shower and underlying event models.
The hadronization correction is obtained from the ratio of $\Rdr$
at the parton level (after the parton shower) and the particle level 
(including all stable particles), both without underlying event.
The underlying event correction is computed from the ratio of $\Rdr$
computed at the particle level with and without underlying event.
The total corrections are defined as the combination of the corrections 
due to hadronization and the underlying event and 
they vary between $+10\%$ and $-3\%$ for tune AMBT1
and between $-1\%$ and $-10\%$ for tune DW.
The results obtained with the two tunes agree typically within
(2--4)\% and always within 11\%.
The central results are taken to be the average values, and the 
uncertainty is taken to be half of the difference 
(given in the supplementary material).
As a cross-check, the non-perturbative corrections have also been
derived with \herwig\ 6.520~\cite{Corcella:2000bw,Corcella:2002jc}.
The \herwig\ results are consistent with the results from the 
\pythia\ tunes AMBT1 and DW for all kinematic regions considered
in this analysis.

The NLO pQCD prediction is given by the ratio of 
an inclusive three-jet cross section and the inclusive jet cross section
both evaluated at their respective NLO.
The numerator and the denominator both depend on the PDFs and
most of the PDF dependencies cancel in the ratio.
A residual PDF dependence remains, due to small differences in the 
decomposition of the partonic subprocesses and a slightly different 
coverage of proton momentum fractions $x$ in the numerator and the 
denominator.
While the PDFs have no explicit $\as$ dependence, their knowledge
(i.e.\ PDF parameterizations) depends implicitly on $\as$
due to assumptions on $\as$ during the extraction procedure.
Therefore, the pQCD prediction for $\Rdr$ has an explicit $\as$ dependence
stemming from the ratios of three-jet and inclusive jet matrix elements,
and an implicit $\as$ dependence due to the residual dependence on the PDFs.

The NLO pQCD results are computed using {\sc fastnlo}~\cite{Kluge:2006xs}
based on {\sc nlojet++}~\cite{Nagy:2003tz,Nagy:2001fj}, in the 
$\overline{\mbox{MS}}$ scheme~\cite{Bardeen:1978yd} for five active 
quark flavors.
The calculations use the next-to-leading logarithmic (two-loop) 
approximation of the RGE and $\as(M_Z)=0.118$ in the matrix elements 
and the PDFs, which is close to the current world average value
of 0.1184~\cite{PDG2012}.
The central choice $\mu_0$ for the renormalization and factorization scales 
is the inclusive jet $p_T$, $\mur = \muf = \mu_0 = p_T$, and the 
MSTW2008NLO PDFs~\cite{Martin:2009iq} are used.

The uncertainties of the pQCD calculations due to uncalculated 
higher order contributions are estimated from the $\mu_{R,F}$ dependence.
These are computed as the relative changes of the results due to 
independent variations of both scales between $\mu_0 / 2 $ and $2\mu_0$, 
with the restriction of $0.5 \le \mur / \muf \le 2.0$.
These variations affect the theory results by (3--9)\%.
The PDF uncertainties are computed using the up and down variations 
of the 20 orthogonal PDF uncertainty eigenvectors,
corresponding to the 68\% C.L., as provided by MSTW2008NLO.
The $\Rdr$ results obtained with the CT10~\cite{Lai:2010vv} 
and NNPDFv2.1~\cite{Ball:2011mu} PDF parameterizations
agree with those for MSTW2008NLO typically within
1\% and always within 3\%.


The theory results are compared to the data in Fig.~\ref{fig:fig1},
and the ratios of data and theory are displayed 
in Fig.~\ref{fig:fig2} for all twelve kinematic regions in $\DR$
and $\ptnbrmin$.
The PDF uncertainties are (2--5)\% and the scale uncertainties are 
typically (4--8)\%.
For higher $\ptnbrmin = 50$, 70, and 90\,GeV, the theoretical predictions 
are in good agreement 
with data and the ratios are independent of $p_T$, $\DR$, and $\ptnbrmin$.
Only for $\ptnbrmin = 30\,$GeV, the predictions are systematically
below the data by (8--15)\%.
This might be caused by limitations of either the perturbative calculation 
or the modeling of the non-perturbative effects at low $\ptnbrmin$.

\begin{table}
\centering
\caption{\label{tab:consistency}
The $\asmz$ results with their absolute uncertainties 
and the $\chi^2$ values from the fits to the $\Rdr$ data
in each of the 12 kinematic regions, defined by the $\ptnbrmin$ and $\DR$
requirements.
}
\begin{ruledtabular}
\begin{tabular}{cccrrr}
$\ptnbrmin$ & $\DR$ & $\asmz$ & \multicolumn{2}{c}{Total uncertainty} & 
    $\chi^2$/$N_{\rm dof}$ \\
\hline
30\,GeV & 1.4--1.8 &  0.1290 & ${+ 0.0073}$ & ${- 0.0078}$   &  6.9 / 11\\
30\,GeV & 1.8--2.2 &  0.1276 & ${+ 0.0078}$ & ${- 0.0049}$  & 12.6 / 11 \\
30\,GeV & 2.2--2.6 & 0.1249 & ${+0.0133}$ & ${- 0.0020}$    & 15.3 / 11 \\
50\,GeV & 1.4--1.8 & 0.1197 & ${+ 0.0089}$ & ${- 0.0061}$   & 7.3 / 11 \\
50\,GeV & 1.8--2.2 & 0.1168 & ${+ 0.0083}$ & ${- 0.0039}$   &  14.1 / 11\\
50\,GeV & 2.2--2.6 & 0.1193 & ${+ 0.0076}$ & ${- 0.0043}$    & 13.7 / 11 \\
70\,GeV & 1.4--1.8 &  0.1168 & ${+0.0101}$ & ${- 0.0073}$   & 4.9 / \phantom{1}9 \\
70\,GeV & 1.8--2.2 & 0.1132 & ${+ 0.0069}$ & ${- 0.0047}$    & 12.1 / 11 \\
70\,GeV & 2.2--2.6 &  0.1156 & ${+ 0.0080}$ & ${- 0.0039}$   & 16.8 / 11 \\
90\,GeV & 1.4--1.8 &  0.1135 & ${+ 0.0084}$ & ${- 0.0087}$   & 1.2 / \phantom{1}9 \\
90\,GeV & 1.8--2.2 &  0.1136 & ${+ 0.0067}$ & ${- 0.0069}$   & 9.7 / \phantom{1}9 \\
90\,GeV & 2.2--2.6 & 0.1166 & ${+ 0.0099}$ & ${- 0.0083}$    & 17.3 / 11\\
\end{tabular}
\end{ruledtabular}
\end{table}

\begin{table*}
\caption{\label{tab:results}
Central values and uncertainties due to different sources
for the 12 $\aspt$ results obtained 
by combining the data at the same $p_T$ from all $\DR$ regions
for $\ptnbrmin = 50$, 70, and 90\,GeV.
All uncertainties are multiplied by a factor of $10^3$.}
\begin{ruledtabular}
\begin{tabular}{cccccccccc}
$p_T$ range  & 
$p_T$  & $\aspt$  &
Total  &
Statistical & Experimental & Non-perturbative & MSTW2008NLO & PDF & $\mu_{R,F}$ \\
(GeV)   &  (GeV)  & & uncertainty & & correlated &
corrections & uncertainty & set & variation \\
\hline
\phantom{1}50 - \phantom{1}60 &
   55.0 &  0.1353 & $^{+ 7.2}_{- 5.6}$ & $\pm 2.8$
   & $^{+ 2.6}_{- 2.8}$ & $^{+ 2.5}_{- 2.8}$ & $^{+ 1.3}_{- 1.2}$ & $^{+ 0.2}_{- 0.4}$
   & $^{+ 5.4}_{- 0.8}$ \\
\phantom{1}60 - \phantom{1}70 &
   65.0 &  0.1299 & $^{+ 8.1}_{- 6.6}$ & $\pm 4.2$
   & $^{+ 2.3}_{- 2.7}$ & $^{+ 2.1}_{- 2.4}$ & $^{+ 1.2}_{- 1.4}$  & $^{+ 0.3}_{- 1.3}$
& $^{+ 6.1}_{- 1.5}$ \\
\phantom{1}70 - \phantom{1}85 &
   77.5 &  0.1232 & $^{+ 4.9}_{- 5.3}$ & $\pm 0.6$
   & $^{+ 1.6}_{- 3.2}$ & $^{+ 1.4}_{-1.0}$ & $^{+ 1.9}_{-1.0}$  & $^{+ 1.8}_{- 0.9}$
& $^{+ 3.5}_{- 3.9}$ \\
\phantom{1}85 - 100 &
   92.5 &  0.1180 & $^{+ 4.9}_{- 3.8}$ & $\pm 0.8$
   & $^{+ 2.8}_{- 2.4}$ & $^{+1.0}_{- 2.2}$ & $^{+ 2.1}_{-1.1}$  & $^{+ 1.0}_{- 0.0}$
& $^{+ 3.0}_{- 1.4}$ \\
100 - 120 &
  110 &  0.1154 & $^{+ 2.8}_{- 7.4}$ & $\pm 0.6$
   & $^{+ 2.1}_{- 2.4}$ & $^{+ 0.3}_{-0.4}$ & $^{+1.0}_{- 5.0}$  & $^{+ 0.0}_{- 3.7}$
& $^{+ 1.4}_{- 3.1}$ \\
120 - 140 &
  130 &  0.1107 & $^{+ 6.0}_{- 3.9}$ & $\pm 0.6$
   & $^{+ 2.8}_{- 2.2}$ & $^{+ 0.4}_{- 0.4}$ & $^{+1.5}_{- 2.5}$  & $^{+ 2.0}_{- 0.0}$
& $^{+ 4.7}_{- 1.9}$ \\
140 - 170 &
  155 &  0.1070 & $^{+ 5.4}_{- 3.8}$ & $\pm 0.5$
   & $^{+ 1.6}_{- 3.0}$ & $^{+ 0.1}_{- 0.3}$ & $^{+ 0.9}_{-0.8}$  & $^{+ 1.5}_{- 0.0}$
& $^{+ 4.9}_{- 2.2}$ \\
170 - 200 &
  185 &  0.1041 & $^{+ 6.7}_{- 4.0}$ & $\pm 0.5$
   & $^{+ 2.5}_{- 2.1}$ & $^{+ 0.7}_{- 0.4}$ & $^{+0.3}_{- 1.5}$  & $^{+ 3.0}_{- 0.0}$
& $^{+ 5.4}_{- 2.9}$ \\
200 - 240 &
  220 &  0.1050 & $^{+ 5.4}_{- 3.3}$ & $\pm 0.3$
   & $^{+ 2.5}_{- 2.3}$ & $^{+ 0.6}_{-0.3}$ & $^{+ 1.0}_{-0.2}$  & $^{+ 0.8}_{- 0.6}$
& $^{+ 4.5}_{- 2.3}$ \\
240 - 280 &
  260 &  0.1061 & $^{+ 5.5}_{- 6.3}$ & $\pm 0.6$
   & $^{+ 1.0}_{- 3.2}$ & $^{+ 1.0}_{-0.8}$ & $^{+ 0.3}_{- 0.7}$  & $^{+ 0.0}_{- 3.3}$
& $^{+ 5.3}_{- 4.2}$ \\
280 - 340 &
  310 &  0.1049 & $^{+ 5.4}_{- 6.2}$ & $\pm 1.0$
   & $^{+ 1.6}_{- 2.3}$ & $^{+ 0.4}_{- 0.3}$ & $^{+ 0.3}_{- 0.6}$  & $^{+ 0.6}_{- 3.3}$
& $^{+ 5.0}_{- 4.3}$ \\
340 - 450 &
  395 &  0.0966 & $^{+ \phantom{1}7.8}_{-10.8}$ & $\pm 5.4$
   & $^{+1.9}_{- 5.9}$ & $^{+ 0.1}_{- 1.0}$ & $^{+ 0.2}_{- 0.9}$  & $^{+ 0.0}_{- 3.3}$
& $^{+ 5.3}_{- 4.7}$ \\
\end{tabular}
\end{ruledtabular}
\end{table*}


These $\Rdr$ results are then used to determine $\as$ and to test the 
two-loop RGE prediction for its running as a function of the scale $p_T$.
In an initial study, the data are split into 12 subsets defined by the 
different ($\DR$, $\ptnbrmin$) requirements.
Assuming the RGE, the value of $\asmz$ is fitted to each of these subsets, 
and the corresponding $\chi^2$ values are determined that compare data 
and theory.
Since each of these subsets covers a large inclusive jet $p_T$ range,
a violation of the RGE would be reflected in poor $\chi^2$ values.
Furthermore, the comparison of the extracted $\asmz$ values allows
the study of the dependence of the results on $\DR$ and/or $\ptnbrmin$.
The data from kinematic regions in ($\DR$, $\ptnbrmin$) in which 
the $\asmz$ fit results are consistent with each other are then used 
in the subsequent analysis.
These data are split into 12 groups, each with the same inclusive jet $p_T$, 
combining data points for different ($\DR$, $\ptnbrmin$).
For each group, $\as$ is determined at the corresponding $p_T$, and then 
evolved, using the RGE, to $\mur = M_Z$.

The $\as$ extraction requires the theory predictions to be available as a 
continuous function of $\as$ used in the matrix elements and PDFs. 
The global PDF fits~\cite{Martin:2009iq,Lai:2010vv,Ball:2011mu} 
do not provide the full $\as$ dependence of their results, but only 
PDF sets at discrete values of $\asmz$, in increments of 
$\Delta \asmz = 0.001$.
A continuous $\asmz$ dependence for $\Rdr$ is obtained, 
by cubic interpolation (linear extrapolation) of the 
theory results inside (outside) the available $\asmz$ range.
For the central results, we use MSTW2008NLO PDFs which cover 
the largest range of $0.110 \le \asmz \le 0.130$.
The fits determine $\as$ by using {\sc minuit}~\cite{minuit} 
to minimize the $\chi^2$ function~\cite{Alekhin:2005dx}
calculated from the differences between theory and data.
All correlated systematic experimental and theoretical uncertainties 
are treated
in the Hessian approach~\cite{Alekhin:2005dx}, except for the 
uncertainty due to the $\mu_{R,F}$ dependence.
The correlated statistical uncertainties are taken into account 
via the covariance matrix.
The $\as$ results are obtained by minimizing $\chi^2$ with respect to $\as$
and the nuisance parameters for the correlated uncertainties.
By scanning $\chi^2$ as a function of $\as$, the uncertainties are 
obtained from those $\as$ values for which $\chi^2$ is increased by one 
with respect to the minimum value.
Fits, that determine $\asmz$ use the two-loop solution of the RGE 
to translate $\asmz$ values to the corresponding values of $\aspt$
which enter the pQCD calculations for the different $p_T$ bins.
These $\asmz$ results are therefore derived assuming the validity of the RGE.
Those fits that extract $\aspt$ from a group of data points in the same 
$p_T$ bin are almost independent of the RGE.
A small dependence on the RGE enters only due to the 
residual dependence of the $\Rdr$ predictions 
on the PDFs which use the RGE in their DGLAP evolution.
Otherwise these $\aspt$ fit results are independent of the RGE.


In the $\as$ determination, we consider the correlations of the 
statistical uncertainties and all 69 sources of correlated experimental 
systematic uncertainties.
The theory uncertainties include the uncertainties of the non-perturbative 
corrections, the PDF uncertainties and the $\mu_{R,F}$ dependence of the 
pQCD calculations.
Following Refs.~\cite{Abazov:2009nc,:2007pb,Chekanov:2006yc}, 
the uncertainty due to the $\mu_{R,F}$ dependence is computed by 
repeating the $\as$ fit for different choices of $\mu_{R,F}$ and 
the largest difference to the central result (obtained for 
$\mu_{R,F} = p_T$) is taken to be the corresponding uncertainty 
for $\as$. 
The $\as$ fits are also repeated for CT10 and NNPDFv2.1 PDFs, and 
the largest differences are quoted as ``PDF set'' uncertainty.
The uncertainties from the scale variation and from the different 
PDF sets are added in quadrature to the other uncertainties to obtain 
the total uncertainty.


Before the central $\as$ results are obtained, the consistency of the 
individual results for the 12 different $(\DR, \ptnbrmin$) regions, 
listed in Table~\ref{tab:consistency}, is tested.
Assuming the RGE, the values of $\asmz$ are fitted to each of the 
12 subsets, and listed in Table~\ref{tab:consistency} together with 
the corresponding $\chi^2$ values.
All $\chi^2$ values are consistent with the expectations based on the 
number of degrees of freedom (N$_{\rm dof}$),
$\chi^2 = {\rm N_{\rm dof}} \pm \sqrt{2 \,{\rm N_{\rm dof}}}$.
This means that the RGE is consistent with the observed $p_T$ dependence 
of $\aspt$ over the studied $p_T$ range in all $\DR$ regions and for 
all $\ptnbrmin$.
For the same $\ptnbrmin$, the $\asmz$ results for different $\DR$ regions 
are consistent with each other, i.e.\ there is no $\DR$ dependence.
The $\asmz$ results are rather independent of $\ptnbrmin$ for 
$\ptnbrmin \ge 50\,$GeV.
Only the $\asmz$ results for the lowest requirement, $\ptnbrmin = 30\,$GeV,
are significantly higher.
As mentioned earlier, at lowest $\ptnbrmin$ limitations of the 
perturbative calculations or the non-perturbative models may 
become visible.
The data with $\ptnbrmin = 30\,$GeV are therefore excluded when 
the final results of this analysis are determined.

\begin{figure}
\includegraphics[scale=1]{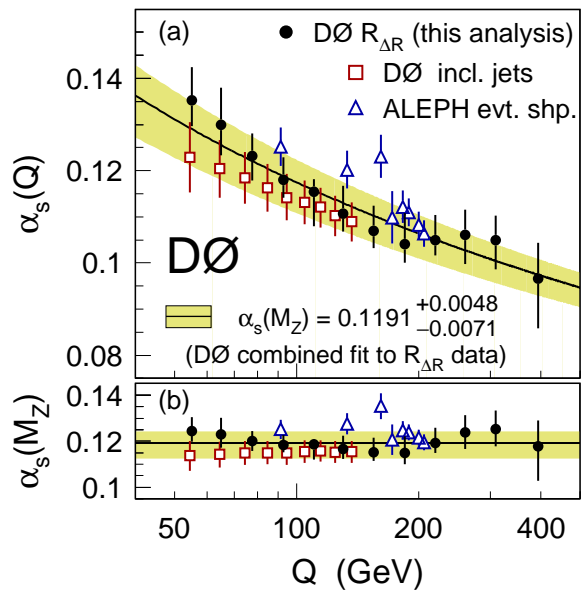}
\caption{(Color online.) 
The strong coupling $\as$ at large momentum transfers, $Q$,
presented as $\asq$ (a) and evolved to $M_Z$ using the RGE (b).
The uncertainty bars indicate the total uncertainty, including
the experimental and theoretical contributions.
The new $\as$ results from $\Rdr$ are compared to previous 
results obtained 
from inclusive jet cross section data~\cite{Abazov:2009nc}
and from event shape data~\cite{Dissertori:2009ik}.
The $\asmz$ result from the combined fit to all selected data points (b)
and the corresponding RGE prediction (a) are also shown.
\label{fig:fig3}}
\end{figure}

All remaining data points with the same $p_T$ 
(from all three $\DR$ regions and for $\ptnbrmin =$ 50, 70, and 90\,GeV)
are combined to fit $\aspt$, at the $p_T$ value corresponding 
to the geometric center of the bin.
This is done for all 12 different $p_T$ bins in the range $50<p_T<450\,$GeV
and the results are listed in Table~\ref{tab:results}
and displayed in Fig.~\ref{fig:fig3} (a).
Using the RGE, the individual results are then evolved to $\mur=M_Z$, 
and shown in Fig.~\ref{fig:fig3} (b).
These $\as$ results from $\Rdr$, extracted using NLO pQCD, are in good 
agreement with our previous results from inclusive jet cross section 
data~\cite{Abazov:2009nc}, extracted using NLO plus 2-loop contributions 
from threshold corrections~\cite{Kidonakis:2000gi}, and with the results 
from a reanalysis of event shape data from the ALEPH experiment at the 
LEP $\epem$ collider, extracted using 
NNLO calculations~\cite{Dissertori:2009ik}. 
A combined fit, using the same data set integrated over $p_T$, and for 
MSTW2008NLO PDFs, gives the $\asmz$ result listed in Table~\ref{tab:asmz}.
The results obtained for CT10 PDFs ($\asmz = 0.1189$) and NNPDFv2.1 
($\asmz = 0.1167$) are used to define the uncertainty due to the PDF set.
This result is in good agreement with our previous result of 
$\asmz = 0.1161 ^{+0.0041}_{-0.0048}$, obtained from inclusive jet cross 
section data at $p_T < 145\,$GeV~\cite{Affolder:2001hn}, and the world 
average value~\cite{PDG2012}.
The RGE prediction for this result is displayed in Fig.~\ref{fig:fig3} (a).
The new $\aspt$ results from $\Rdr$ are well described by the RGE prediction
including the region $208 < \mur < 400\,$GeV, in which the RGE is tested
for the first time.

\begin{table*}
\centering
\caption{\label{tab:asmz}
The $\asmz$ result for $\Rdr$, obtained by combining all data points
in $p_T$ and in $\DR$ for the requirements $\ptnbrmin = 50$, 70, and 90\,GeV.
All uncertainties are multiplied by a factor of $10^3$.}
\begin{ruledtabular}
\begin{tabular}{cccccccc}
$\asmz$  &
Total  &
Statistical  &   Experimental  &    Non-perturbative  & MSTW2008NLO   & PDF & $\mu_{R,F}$ \\
    & uncertainty &   & correlated &
corrections & uncertainty &  set & variation \\
\hline
  0.1191 & $^{+ 4.8}_{- 7.1}$ & $\pm 0.3$
 & $^{+ 0.7}_{- 0.9}$ & $^{+ 0.2}_{-0.1}$ & $^{+ 1.0}_{- 0.5}$ & $^{+ 0.0}_{- 2.4}$
& $^{+ 4.6}_{- 6.6}$ \\
\end{tabular}
\end{ruledtabular}
\end{table*}


In summary, a measurement has been presented of a new quantity $\Rdr$
which probes the angular correlations of jets.
$\Rdr$ is measured as a function of inclusive jet $p_T$ in 
different annular regions of $\DR$ between a jet and its neighboring jets
and for different requirements on the minimal transverse momentum 
of the neighboring jet $\ptnbrmin$.
The data for $p_T > 50\,$GeV are well-described by pQCD calculations 
in NLO in $\as$ with non-perturbative corrections applied.
Results for $\aspt$ are extracted using the data with $\ptnbrmin \ge 50\,$GeV,
integrated over $\DR$.
The extracted $\aspt$ results from $\Rdr$ are, to good approximation, 
independent of the PDFs and thus independent of assumptions on the RGE.
Therefore, these $\as$ results are the first to provide a test of 
the RGE at momentum transfers beyond 208\,GeV.
The results are in good agreement with previous results and consistent with 
the RGE predictions  for the running of $\as$ for momentum transfers 
up to 400\,GeV.
The combined $\asmz$ result, obtained using the data with 
$\ptnbrmin \ge 50\,$GeV (integrated over $\DR$ and $p_T$), 
is $\asmz = 0.1191^{+0.0048}_{-0.0071}$, in good agreement
with the world average value~\cite{PDG2012}.

\input acknowledgement.tex
\appendix
\section{Appendix A. Supplementary Material}
Supplemantary material related to this letter can be found
online at http://dx.doi.org/10.1016/j.physletb.2012.10.003.

\end{document}

%% file: author_list.tex
\affiliation{LAFEX, Centro Brasileiro de Pesquisas F\'{i}sicas, Rio de Janeiro, Brazil}
\affiliation{Universidade do Estado do Rio de Janeiro, Rio de Janeiro, Brazil}
\affiliation{Universidade Federal do ABC, Santo Andr\'e, Brazil}
\affiliation{University of Science and Technology of China, Hefei, People's Republic of China}
\affiliation{Universidad de los Andes, Bogot\'a, Colombia}
\affiliation{Charles University, Faculty of Mathematics and Physics, Center for Particle Physics, Prague, Czech Republic}
\affiliation{Czech Technical University in Prague, Prague, Czech Republic}
\affiliation{Center for Particle Physics, Institute of Physics, Academy of Sciences of the Czech Republic, Prague, Czech Republic}
\affiliation{Universidad San Francisco de Quito, Quito, Ecuador}
\affiliation{LPC, Universit\'e Blaise Pascal, CNRS/IN2P3, Clermont, France}
\affiliation{LPSC, Universit\'e Joseph Fourier Grenoble 1, CNRS/IN2P3, Institut National Polytechnique de Grenoble, Grenoble, France}
\affiliation{CPPM, Aix-Marseille Universit\'e, CNRS/IN2P3, Marseille, France}
\affiliation{LAL, Universit\'e Paris-Sud, CNRS/IN2P3, Orsay, France}
\affiliation{LPNHE, Universit\'es Paris VI and VII, CNRS/IN2P3, Paris, France}
\affiliation{CEA, Irfu, SPP, Saclay, France}
\affiliation{IPHC, Universit\'e de Strasbourg, CNRS/IN2P3, Strasbourg, France}
\affiliation{IPNL, Universit\'e Lyon 1, CNRS/IN2P3, Villeurbanne, France and Universit\'e de Lyon, Lyon, France}
\affiliation{III. Physikalisches Institut A, RWTH Aachen University, Aachen, Germany}
\affiliation{Physikalisches Institut, Universit\"at Freiburg, Freiburg, Germany}
\affiliation{II. Physikalisches Institut, Georg-August-Universit\"at G\"ottingen, G\"ottingen, Germany}
\affiliation{Institut f\"ur Physik, Universit\"at Mainz, Mainz, Germany}
\affiliation{Ludwig-Maximilians-Universit\"at M\"unchen, M\"unchen, Germany}
\affiliation{Fachbereich Physik, Bergische Universit\"at Wuppertal, Wuppertal, Germany}
\affiliation{Panjab University, Chandigarh, India}
\affiliation{Delhi University, Delhi, India}
\affiliation{Tata Institute of Fundamental Research, Mumbai, India}
\affiliation{University College Dublin, Dublin, Ireland}
\affiliation{Korea Detector Laboratory, Korea University, Seoul, Korea}
\affiliation{CINVESTAV, Mexico City, Mexico}
\affiliation{Nikhef, Science Park, Amsterdam, the Netherlands}
\affiliation{Radboud University Nijmegen, Nijmegen, the Netherlands}
\affiliation{Joint Institute for Nuclear Research, Dubna, Russia}
\affiliation{Institute for Theoretical and Experimental Physics, Moscow, Russia}
\affiliation{Moscow State University, Moscow, Russia}
\affiliation{Institute for High Energy Physics, Protvino, Russia}
\affiliation{Petersburg Nuclear Physics Institute, St. Petersburg, Russia}
\affiliation{Instituci\'{o} Catalana de Recerca i Estudis Avan\c{c}ats (ICREA) and Institut de F\'{i}sica d'Altes Energies (IFAE), Barcelona, Spain}
\affiliation{Uppsala University, Uppsala, Sweden}
\affiliation{Lancaster University, Lancaster LA1 4YB, United Kingdom}
\affiliation{Imperial College London, London SW7 2AZ, United Kingdom}
\affiliation{The University of Manchester, Manchester M13 9PL, United Kingdom}
\affiliation{University of Arizona, Tucson, Arizona 85721, USA}
\affiliation{University of California Riverside, Riverside, California 92521, USA}
\affiliation{Florida State University, Tallahassee, Florida 32306, USA}
\affiliation{Fermi National Accelerator Laboratory, Batavia, Illinois 60510, USA}
\affiliation{University of Illinois at Chicago, Chicago, Illinois 60607, USA}
\affiliation{Northern Illinois University, DeKalb, Illinois 60115, USA}
\affiliation{Northwestern University, Evanston, Illinois 60208, USA}
\affiliation{Indiana University, Bloomington, Indiana 47405, USA}
\affiliation{Purdue University Calumet, Hammond, Indiana 46323, USA}
\affiliation{University of Notre Dame, Notre Dame, Indiana 46556, USA}
\affiliation{Iowa State University, Ames, Iowa 50011, USA}
\affiliation{University of Kansas, Lawrence, Kansas 66045, USA}
\affiliation{Kansas State University, Manhattan, Kansas 66506, USA}
\affiliation{Louisiana Tech University, Ruston, Louisiana 71272, USA}
\affiliation{Boston University, Boston, Massachusetts 02215, USA}
\affiliation{Northeastern University, Boston, Massachusetts 02115, USA}
\affiliation{University of Michigan, Ann Arbor, Michigan 48109, USA}
\affiliation{Michigan State University, East Lansing, Michigan 48824, USA}
\affiliation{University of Mississippi, University, Mississippi 38677, USA}
\affiliation{University of Nebraska, Lincoln, Nebraska 68588, USA}
\affiliation{Rutgers University, Piscataway, New Jersey 08855, USA}
\affiliation{Princeton University, Princeton, New Jersey 08544, USA}
\affiliation{State University of New York, Buffalo, New York 14260, USA}
\affiliation{University of Rochester, Rochester, New York 14627, USA}
\affiliation{State University of New York, Stony Brook, New York 11794, USA}
\affiliation{Brookhaven National Laboratory, Upton, New York 11973, USA}
\affiliation{Langston University, Langston, Oklahoma 73050, USA}
\affiliation{University of Oklahoma, Norman, Oklahoma 73019, USA}
\affiliation{Oklahoma State University, Stillwater, Oklahoma 74078, USA}
\affiliation{Brown University, Providence, Rhode Island 02912, USA}
\affiliation{University of Texas, Arlington, Texas 76019, USA}
\affiliation{Southern Methodist University, Dallas, Texas 75275, USA}
\affiliation{Rice University, Houston, Texas 77005, USA}
\affiliation{University of Virginia, Charlottesville, Virginia 22901, USA}
\affiliation{University of Washington, Seattle, Washington 98195, USA}
\author{V.M.~Abazov} \affiliation{Joint Institute for Nuclear Research, Dubna, Russia}
\author{B.~Abbott} \affiliation{University of Oklahoma, Norman, Oklahoma 73019, USA}
\author{B.S.~Acharya} \affiliation{Tata Institute of Fundamental Research, Mumbai, India}
\author{M.~Adams} \affiliation{University of Illinois at Chicago, Chicago, Illinois 60607, USA}
\author{T.~Adams} \affiliation{Florida State University, Tallahassee, Florida 32306, USA}
\author{G.D.~Alexeev} \affiliation{Joint Institute for Nuclear Research, Dubna, Russia}
\author{G.~Alkhazov} \affiliation{Petersburg Nuclear Physics Institute, St. Petersburg, Russia}
\author{A.~Alton$^{a}$} \affiliation{University of Michigan, Ann Arbor, Michigan 48109, USA}
\author{G.~Alverson} \affiliation{Northeastern University, Boston, Massachusetts 02115, USA}
\author{A.~Askew} \affiliation{Florida State University, Tallahassee, Florida 32306, USA}
% ...
%
%
\collaboration{The D0 Collaboration\footnote{with visitors from
$^{a}$Augustana College, Sioux Falls, SD, USA,
$^{b}$The University of Liverpool, Liverpool, UK,
$^{c}$UPIITA-IPN, Mexico City, Mexico,
$^{d}$DESY, Hamburg, Germany,
,
$^{e}$SLAC, Menlo Park, CA, USA,
$^{f}$University College London, London, UK,
$^{g}$Centro de Investigacion en Computacion - IPN, Mexico City, Mexico,
$^{h}$ECFM, Universidad Autonoma de Sinaloa, Culiac\'an, Mexico
and
$^{i}$Universidade Estadual Paulista, S\~ao Paulo, Brazil.
}} \noaffiliation
\vskip 0.25cm

%% file: acknowledgement.tex
%
We thank the staffs at Fermilab and collaborating institutions,
and acknowledge support from the
DOE and NSF (USA);
CEA and CNRS/IN2P3 (France);
MON, NRC KI and RFBR (Russia);
CNPq, FAPERJ, FAPESP and FUNDUNESP (Brazil);
DAE and DST (India);
Colciencias (Colombia);
CONACyT (Mexico);
NRF (Korea);
FOM (The Netherlands);
STFC and the Royal Society (United Kingdom);
MSMT and GACR (Czech Republic);
BMBF and DFG (Germany);
SFI (Ireland);
The Swedish Research Council (Sweden);
and
CAS and CNSF (China).

%% file: rdr.bbl
\begin{thebibliography}{99}
\bibitem{rapidity}
  Rapidity $y$ is related to the polar scattering angle $\theta$
  with respect to the proton beam axis by 
  $y=\frac{1}{2} \ln \left[ (1+\beta\cos\theta)/(1-\beta \cos \theta) \right]$,
  where $\beta$ is defined as the ratio of
  the magnitude of the momentum and energy,
  $\beta=|\vec{p}| / E$.
\end{thebibliography}
